\documentstyle[12pt]{article}
\begin{document}

\begin{flushright}
PITHA 94/37\\
hep-ph/9409396\\
September, 1994
\end{flushright}
\vspace*{2cm}
\LARGE\centerline{Energy Spectra and Energy Correlations}
\centerline{in the Decay $H\to ZZ\to \mu^+\mu^-\mu^+\mu^-$}
\vspace*{1cm}
\large\centerline{Torsten Arens and L. M. Sehgal}
\centerline{III. Physikalisches Institut (A), RWTH Aachen,}
\centerline{D-52074 Aachen, Germany}
\normalsize

\begin{abstract}
It is shown that in the sequential decay $H\to ZZ\to (f_1\bar{f_1})+
(f_2\bar{f_2})$, the energy distribution of the final state particles
provides a simple and powerful test of the $HZZ$ vertex. For a standard
Higgs boson, the energy spectrum of any final fermion, in the rest frame
of $H$, is predicted to be $d\Gamma /dx\sim 1+\beta^4-2(x-1)^2$, with
$\beta = \sqrt{1-4m^2_Z/m^2_H}$ and $1-\beta \le x=4E/m_H\le 1+\beta $.
By contrast, the spectrum for a pseudoscalar Higgs is $d\Gamma /dx
\sim \beta^2+(x-1)^2$. There are characteristic energy correlations
between $f_1$ and $f_2$ and between $f_1$ and $\bar{f_2}$. These
considerations are applied to the ``gold--plated'' reaction $H\to ZZ\to
\mu^+\mu^-\mu^+\mu^-$, including possible effects of $CP$--violation in
the $HZZ$ coupling. Our formalism also yields the energy spectra and
correlations of leptons in the decay $H\to W^+W^-\to l^+\nu_ll^-
\bar{\nu_l}$.
\end{abstract}

\newpage
\section{Introduction}
One of the distinctive signatures of a Higgs particle with mass $m_H>
2m_Z$ is the sequential decay
\begin{eqnarray}
H\to ZZ\to (\mu^+\mu^-)+(\mu^+\mu^-).
\end{eqnarray}
The observation of such a four--muon final state, consisting of two
$\mu^+\mu^-$ pairs with the invariant mass of the $Z$, together
resonating at some invariant mass $m_H$, would be unmistakable evidence
for a particle with the \it prima facie \rm characteristics of the
Higgs boson \cite{higgs}.\\
\indent In the standard model, the Higgs boson has the quantum numbers
$J^{PC}=0^{++}$, and a very specific form of coupling to gauge bosons,
namely $g_{\mu\nu}\varepsilon_1^{\mu}\varepsilon_2^{\nu}$. (We denote
the polarization vectors and momenta of the two $Z$ bosons by
($\varepsilon_1,p_1$) and ($\varepsilon_2,p_2$).) More generally, a
scalar $0^{++}$ particle can couple to a pair of $Z$'s according to
$(Bg_{\mu\nu}+\displaystyle\frac{C}{m_Z^2}p_{1\nu}p_{2\mu})
\varepsilon_1^{\mu}\varepsilon_2^{\nu}$. In extended gauge models,
there are also pseudoscalar Higgs particles with quantum numbers $0^{-+}$;
these can have radiatively induced couplings to two $Z$ bosons of the
form $\displaystyle\frac{D}{m^2_Z}\varepsilon_{\mu\nu\rho\sigma}
\varepsilon_1^{\mu}\varepsilon_2^{\nu}p_1^{\rho}p_2^{\sigma}$
\cite{nelson, soni, chang1}.\\
\indent The purpose of this paper is to show that in decays of the type
\begin{eqnarray}
H\to ZZ\to (f_1\bar{f_1})+(f_2\bar{f_2}),
\end{eqnarray}
of which Eq. (1) is an example, the energy spectrum of the final fermions, in
the rest frame of the $H$, is a simple and powerful probe of the $HZZ$ vertex.
This is highlighted by the following result (to be obtained in Section 3): For
a standard Higgs boson, the energy spectrum of any final fermion (independent
of whether it is a lepton, $u$--quark or $d$--quark) has the universal form
\begin{eqnarray}
\frac{1}{\Gamma}\frac{\mbox{d}\Gamma}{\mbox{d}x}=\frac{3/(2\beta )}
{3-2\beta^2+3\beta^4}
\Bigl [1+\beta^4-2(x-1)^2\Bigr ]\qquad\mbox{(Scalar Higgs)},
\end{eqnarray}
where $x=4E/m_H$, $\beta =\sqrt{1-4m_Z^2/m_H^2}$ and the range of $x$ is
$1-\beta\le x\le 1+\beta$. By contrast, for a pseudoscalar Higgs boson, the
spectrum is
\begin{eqnarray}
\frac{1}{\Gamma}\frac{\mbox{d}\Gamma}{\mbox{d}x}=\frac{3}{8\beta^3}
\Bigl [\beta^2+(x-1)^2\Bigr ]\qquad\mbox{(Pseudoscalar Higgs)}.
\end{eqnarray}
This difference between $0^{++}$ and $0^{-+}$ Higgs decays is exhibited in
Fig. 1, for $m_H=300$ GeV.\\
\indent We will derive in this paper the energy spectrum of the reaction (2)
for a general $HZZ$ coupling. In addition to single particle spectra, we will
obtain the correlated two--particle energy distribution of $f_1$ and $f_2$,
 and of $f_1$ and $\bar{f_2}$. In Section 3, we consider Higgs couplings of
the scalar and pseudoscalar form, as well as $CP$--violating effects when
both are present simultaneously. In Section 4, we will relate the energy
characteristics of the reactions (1) and (2) to the helicity wave--function
of the $ZZ$ system created in the decay $H\to ZZ$.\\
\indent Our work is complementary to other analyses devoted to the sequential
decays $H\to ZZ\to (f_1\bar{f_1})+(f_2\bar{f_2})$, in which the structure of
the $HZZ$ coupling is probed by means of the angular distribution of the final
particles, particularly the correlation between the $f_1\bar{f_1}$ and $f_2
\bar{f_2}$ planes \cite{nelson} - \cite{djouadi}.
Our formalism allows us also to obtain the energy spectrum
and correlation of leptons produced in the decay $H\to W^+W^-\to l^+\nu_ll^-
\bar{\nu_l}$, which we have reported earlier \cite{arens}.

\section{Differential Decay Rate}
Consider the sequential decay
\begin{eqnarray}
H(P)\to Z(p_1)Z(p_2)\to f_1(q_1)\bar{f_1}(q_2)f_2(q_3)\bar{f_2}(q_4)
\end{eqnarray}
induced by a general $HZZ$ coupling
\begin{eqnarray}
A\Bigl [H\to Z(\varepsilon_1, p_1)+Z(\varepsilon_2, p_2)\Bigr ]=
2im_Z^2\sqrt{G_F\sqrt{2}}\Bigl [Bg_{\mu\nu}+\frac{C}
{m_Z^2}p_{1\nu}p_{2\mu}\nonumber\\
+\frac{D}{m_Z^2}\varepsilon_{\mu\nu\rho\sigma}
p_1^{\rho}p_2^{\sigma}\Bigr ]\varepsilon_1^{*\mu}\varepsilon_2^{*\nu}.
\end{eqnarray}
The differential decay rate is given by
\begin{eqnarray}
\mbox{d}^{8}\Gamma&=&\frac{8\sqrt{2}G_F^3m_Z^4D_Z}{m_H}(v_1^2+a_1^2)
(v_2^2+a_2^2)\biggl [
|B|^2{\cal S}+\frac{|C|^2}{m_Z^4}{\cal L}+\frac{\mbox{Re}(B^*C)}{m_Z^2}
{\cal M}\nonumber\\
&&+\frac{\mbox{Im}(B^*C)}{m_Z^2}{\cal N}+\frac{|D|^2}{m_Z^4}{\cal P}
+\frac{\mbox{Re}(B^*D)}{m_Z^2}{\cal Q}+\frac{\mbox{Im}(B^*D)}{m_Z^2}{\cal R}
\nonumber\\
&&+\frac{\mbox{Re}(C^*D)}{m_Z^4}{\cal U}+\frac{\mbox{Im}(C^*D)}{m_Z^4}{\cal V}
\biggr ]\cdot\mbox{d}Lips ,
\end{eqnarray}
where, neglecting fermion masses,
\begin{eqnarray}
{\cal S}&=&(q_1\cdot q_3)(q_2\cdot q_4)+(q_1\cdot q_4)(q_2\cdot q_3)
+\xi_1\xi_2\Bigl ((q_1\cdot q_3)(q_2\cdot q_4)-(q_1\cdot q_4)
(q_2\cdot q_3)\Bigr ),\nonumber\\
{\cal L}&=&2\Bigl ((p_2\cdot q_1)(p_2\cdot q_2)
-\frac{m_Z^4}{4}\Bigl )\Bigl ((p_1\cdot q_3)(p_1\cdot q_4)
-\frac{m_Z^4}{4}\Bigl ),\nonumber\\
{\cal M}&=&(p_2\cdot q_1)\Bigl ((p_1\cdot q_3)(q_2\cdot q_4)+(p_1\cdot q_4)
(q_2\cdot q_3)\Bigr )\nonumber\\
&+&(p_2\cdot q_2)\Bigl ((p_1\cdot q_3)(q_1\cdot q_4)+
(p_1\cdot q_4)(q_1\cdot q_3)\Bigr )-\frac{m_Z^4}{4}(p_1\cdot p_2)
\nonumber\\
&+&\xi_1\xi_2\biggl [(p_1\cdot p_2)\Bigl ((q_1\cdot q_3)(q_2\cdot q_4)-
(q_1\cdot q_4)(q_2\cdot q_3)\Bigr )\nonumber\\
&&-\frac{m_Z^4}{4}\Bigl ((q_1-q_2)\cdot
(q_3-q_4)\Bigr )\biggr ],\nonumber\\
{\cal N}&=&\varepsilon (q_1,q_2,q_3,q_4)\biggl [\xi_1\Bigl (p_1\cdot
(q_4-q_3)\Bigr )+\xi_2\Bigl (p_2\cdot (q_2-q_1)\Bigr )\biggr ],\nonumber\\
{\cal P}&=&-\frac{m_Z^8}{8}-2\Bigl ((q_1\cdot q_3)(q_2\cdot q_4)-
(q_1\cdot q_4)(q_2\cdot q_3)\Bigl )^2\nonumber\\
&+&\frac{m_Z^4}{4}\biggl [\Bigl ((q_1\cdot q_3)+(q_2\cdot q_4)\Bigr )^2
+\Bigl ((q_1\cdot q_4)+(q_2\cdot q_3)\Bigr )^2
\biggr ]\nonumber\\
&+&\xi_1\xi_2\frac{m_Z^4}{4}\biggl [\Bigl ((q_1\cdot q_3)-
(q_2\cdot q_4)\Bigr )^2-\Bigl ((q_1\cdot q_4)-(q_2\cdot q_3)\Bigr )^2
\biggr ],\nonumber\\
{\cal Q}&=&\varepsilon (q_1,q_2,q_3,q_4)\biggl [
\Bigl ((q_1-q_2)\cdot (q_4-q_3)\Bigr )-\xi_1\xi_2(p_1\cdot p_2)
\biggr ],\nonumber\\
{\cal R}&=&(\xi_1+\xi_2)\biggl [\Bigl ((q_1\cdot q_3)-(q_2\cdot q_4)
\Bigr )\Bigl (\frac{m_Z^4}{4}+(q_1\cdot q_3)(q_2\cdot q_4)-
(q_1\cdot q_4)(q_2\cdot q_3)\Bigr)\biggr ]\nonumber\\
&+&(\xi_1-\xi_2)\biggl [\Bigl ((q_1\cdot q_4)-(q_2\cdot q_3)\Bigr )\Bigl (
\frac{m_Z^4}{4}+(q_1\cdot q_4)(q_2\cdot q_3)-(q_1\cdot q_3)(q_2\cdot q_4)
\Bigr)\biggr ],\nonumber\\
{\cal U}&=&\varepsilon (q_1,q_2,q_3,q_4)\biggl [
\Bigl (p_1\cdot (q_3-q_4)\Bigr )\Bigl (p_2\cdot (q_2-q_1)\Bigr )
-\xi_1\xi_2\Bigl ((p_1\cdot p_2)^2-m_Z^4\Bigr )\biggr ],\nonumber\\
{\cal V}&=&\biggl [\xi_1\Bigl (p_1\cdot (q_3-q_4)\Bigr )
+\xi_2\Bigl (p_2\cdot (q_1-q_2)\Bigr )\biggr ]\biggl [
\frac{m_Z^4}{4}\Bigl ((q_1-q_2)\cdot (q_4-q_3)\Bigl )\nonumber\\
&&+(p_1\cdot p_2)\Bigl ((q_1\cdot q_3)(q_2\cdot q_4)-
(q_1\cdot q_4)(q_2\cdot q_3)\Bigr )\biggr ],
\end{eqnarray}
and $\mbox{d}Lips$ is the Lorentz invariant phase space element
\begin{eqnarray}
\mbox{d}Lips = (2\pi )^4\delta^{(4)}(P-q_1-
q_2-q_3-q_4)\prod_{i=1}^4\frac{\mbox{d}^3q_i}{(2\pi )^32q_i^0}.
\end{eqnarray}
The factor $D_Z$ in Eq. (7) is the product of the two $Z$ boson
propagators,
\begin{eqnarray}
D_Z=m_Z^4\prod^2_{j=1}\frac{1}{(p_j^2-m_Z^2)^2+m_Z^2\Gamma^2_Z},
\end{eqnarray}
which, in the narrow--width approximation, may be written as
\begin{eqnarray}
D_Z\approx \frac{\pi^2m_Z^2}{\Gamma^2_Z}\delta (p_1^2-m_Z^2)\delta (p_2^2
-m_Z^2).
\end{eqnarray}
The parameters $\xi_1$ and $\xi_2$ are given by
\begin{eqnarray}
\xi_i=\frac{2v_ia_i}{v_i^2+a_i^2},\qquad i=1,2
\end{eqnarray}
where $v_i$ and $a_i$ are the vector and axial vector coupling constants of
the fermion pair $f_i\bar{f_i}$ to $Z$ ($v_f=2I^3_f-4e_f\sin^2\Theta_W$,
$a_f=2I^3_f$).
Eq. (7) is the basis of all the results to be obtained in the following
Sections.

\section{Energy Spectra and Correlations: Scalar vs. Pseudoscalar Higgs
and $CP$--Violation}
We consider an $HZZ$ coupling that is a combination of $B$-- and
$D$--type terms in Eq. (6), with $C=0$. This will enable us to compare a
scalar Higgs ($B\neq 0$, $D=0$) with a pseudoscalar Higgs ($B=0$, $D\neq 0$),
as well as discuss certain $CP$--violating effects arising from
simultaneous presence of both terms.\\
\indent For the reaction $H\to ZZ\to (f_1\bar{f_1})+(f_2\bar{f_2})$, with
$f_1\neq f_2$, we obtain the following energy distributions for the pairs
$(f_1,f_2)$, $(\bar{f_1},\bar{f_2})$, $(\bar{f_1},f_2)$ and
$(f_1,\bar{f_2})$:
\begin{eqnarray}
\frac{1}{\Gamma}\frac{\mbox{d}\Gamma}{\mbox{d}x(\stackrel{(-)}{f_1})
\mbox{d}x'(\stackrel{(-)}{f_2})}&=&\frac{1}{N}\biggl\{|B|^2
(F_1+\xi_1\xi_2F_2)+|D|^2(F_3+\xi_1\xi_2F_4)\nonumber\\
&&\stackrel{(-)}{+}\mbox{Im}(B^*D)\Bigl [\xi_1(F_5+F_6)
+\xi_2(F_5-F_6)\Bigr ]\biggr\},\nonumber\\
\frac{1}{\Gamma}\frac{\mbox{d}\Gamma}{\mbox{d}x(\bar{f_1})
\mbox{d}x'(f_2)}&=&\frac{1}{N}
\biggl\{|B|^2(F_1-\xi_1\xi_2F_2)
+|D|^2(F_3-\xi_1\xi_2F_4)\nonumber\\
&&-\mbox{Im}(B^*D)\Bigl [\xi_1(F_5+F_6)
-\xi_2(F_5-F_6)\Bigr ]\biggr\},\nonumber\\
\frac{1}{\Gamma}\frac{\mbox{d}\Gamma}{\mbox{d}x(f_1)
\mbox{d}x'(\bar{f_2})}&=&\frac{1}{N}
\biggl\{|B|^2(F_1-\xi_1\xi_2F_2)
+|D|^2(F_3-\xi_1\xi_2F_4)\nonumber\\
&&+\mbox{Im}(B^*D)\Bigl [\xi_1(F_5+F_6)
-\xi_2(F_5-F_6)\Bigr ]\biggr\},
\end{eqnarray}
with the normalization factor
\begin{eqnarray}
N=|B|^2(3-2\beta^2+3\beta^4)+8|D|^2\beta^2.
\end{eqnarray}
The functions $F_i$ are defined as follows
\begin{eqnarray}
F_1(x,x')&=&\frac{9}{32\beta^6}\biggl\{(1-\beta^2)^2\Bigl [\beta^2+(x-1)^2
\Bigr ]\Bigl [\beta^2+(x'-1)^2\Bigr ]\nonumber\\
&&+2(1+\beta^2)^2\Bigl [\beta^2-(x-1)^2\Bigr ]\Bigl [\beta^2-(x'-1)^2
\Bigr ]\biggr\},\nonumber\\
F_2(x,x')&=&\frac{9}{8\beta^4}(1-\beta^2)^2(x-1)(x'-1),\nonumber\\
F_3(x,x')&=&\frac{9}{8\beta^4}\Bigl [\beta^2+(x-1)^2\Bigr ]\Bigl [\beta^2+
(x'-1)^2\Bigr ],\nonumber\\
F_4(x,x')&=&4\beta^2 F_2(x,x')/(1-\beta^2)^2,\nonumber\\
F_5(x,x')&=&\frac{9(1-\beta^2)}{8\beta^4}(x+x'-2)\Bigl [(x-1)(x'-1)+\beta^2
\Bigr ],\nonumber\\
F_6(x,x')&=&\frac{9(1-\beta^2)}{8\beta^4}(x'-x)\Bigl [(x-1)(x'-1)-\beta^2
\Bigr ].
\end{eqnarray}
Note that $F_{1,\cdots ,5}$ are symmetric under
exchange of $x$ and $x'$, while $F_6$ is antisymmetric. \\
\indent By integrating the two--particle distributions (13) over the
energies of one of the particles, we obtain the inclusive energy spectrum of
any fermion $f$ in the decay $H\to ZZ\to f\bar{f}+\cdots $:
\begin{eqnarray}
\frac{1}{\Gamma}\frac{\mbox{d}\Gamma}{\mbox{d}x(\stackrel{(-)}{f})}&=&
\frac{3}{2\beta N}
\biggl\{|B|^2\Bigl [1+\beta^4-2(x-1)^2\Bigr ]+2|D|^2\Bigl [\beta^2+
(x-1)^2\Bigr ]\nonumber\\
&&\quad\stackrel{(-)}{+}4\mbox{Im}(B^*D)\xi(1-\beta^2)(x-1)\biggr\}.
\end{eqnarray}
For a pure scalar ($B$--type) or pseudoscalar ($D$--type) Higgs coupling, we
obtain the results already announced in Eqs. (3) and (4).\\
\indent Although the two--particle distributions given in Eq. (13) were
derived for the reaction $H\to ZZ\to (f_1\bar{f_1})+(f_2\bar{f_2})$ with
$f_1\neq f_2$, the results are also applicable to the decay $H\to ZZ\to
(\mu^+\mu^-)+(\mu^+\mu^-)$. The only requirement is that the two observed
muons belong to different $Z$'s. This is automatically the case for like
sign pairs $\mu^+\mu^+$ or $\mu^-\mu^-$, and is easy to ensure for $\mu^+
\mu^-$ by requiring that their invariant mass be different from $m_Z$.
(Note that a $\mu^+\mu^-$ pair from the same $Z$ must fulfil $x+x'=2$.)
With this proviso the correlated energy distributions for $\mu^+\mu^+$,
$\mu^-\mu^-$ and $\mu^+\mu^-$ are
\begin{eqnarray}
\frac{1}{\Gamma}\frac{\mbox{d}\Gamma}{\mbox{d}x(\mu^{\pm})
\mbox{d}x'(\mu^{\pm})}\!\!&\!\!=\!\!&\!\!
\frac{1}{N}\Bigl [|B|^2(F_1+\xi^2F_2)+|D|^2(F_3+\xi^2F_4)
\mp 2\mbox{Im}(B^*D)\xi F_5\Bigr ],\nonumber\\
\frac{1}{\Gamma}\frac{\mbox{d}\Gamma}{\mbox{d}x(\mu^{\pm})
\mbox{d}x'(\mu^{\mp})}\!\!&\!\!=\!\!&\!\!
\frac{1}{N}\Bigl [|B|^2(F_1-\xi^2F_2)+|D|^2(F_3-\xi^2F_4)
\mp 2\mbox{Im}(B^*D)\xi F_6\Bigr ].
\end{eqnarray}
Here $\xi =2va/(v^2+a^2)\approx 0.16$ is the parameter describing the $Z$
coupling to muons.\\
\indent From Eqs. (17), we draw the following conclusions.\\
\indent (i) For a scalar Higgs coupling ($D=0$), the like sign muon pairs
$\mu^+\mu^+$ and $\mu^-\mu^-$ have the energy distribution
\begin{eqnarray}
\frac{1}{\Gamma}\frac{\mbox{d}\Gamma}{\mbox{d}x(\mu^{\pm})\mbox{d}
x'(\mu^{\pm})}=
\frac{1}{3-2\beta^2+3\beta^4}\Bigl [F_1(x,x')+\xi^2F_2(x,x')\Bigr ]\qquad
\mbox{(Scalar Higgs)}.
\end{eqnarray}
This is quite distinct from the pseudoscalar Higgs decay, given by
\begin{eqnarray}
\frac{1}{\Gamma}\frac{\mbox{d}\Gamma}{\mbox{d}x(\mu^{\pm})\mbox{d}
x'(\mu^{\pm})}=
\frac{1}{8\beta^2}\Bigl [F_3(x,x')+\xi^2F_4(x,x')\Bigr ]\qquad
\mbox{(Pseudoscalar Higgs)}.
\end{eqnarray}
These two distributions are contrasted in Fig. 2. \\
\indent (ii) The unlike sign dimuon $\mu^+\mu^-$ spectra are likewise
quite different for $0^{++}$ and $0^{-+}$ Higgs decays. Indeed, the $\mu^+
\mu^-$ spectra differ from the $\mu^{\pm}\mu^{\pm}$ given in Eqs. (18) and
(19) only in the replacement $\xi^2\to -\xi^2$. Since $\xi^2$ is
approximately $2.56\times 10^{-2}$, the two--dimensional distributions
for unlike sign $\mu$'s are very similar to those of the like sign muon
pairs, shown in Fig. 2.\\
\indent (iii) The simultaneous presence of $B$-- and $D$--type couplings
produces a $CP$--violating asymmetry between $\mu^+\mu^+$ and $\mu^-\mu^-$.
There is also an asymmetry between the single particle $\mu^+$ and $\mu^-$
spectra, that can be read off Eq. (16):
\begin{eqnarray}
A&=&\frac{\mbox{d}\Gamma /\mbox{d}x(\mu^-)-\mbox{d}\Gamma /\mbox{d}x(\mu^+)}
{\mbox{d}\Gamma /\mbox{d}x(\mu^-)+\mbox{d}\Gamma /\mbox{d}x(\mu^+)}\nonumber\\
&=&\mbox{Im}(B^*D)\frac{4\xi (x-1)(1-\beta^2)}{|B|^2\Bigl [1+\beta^4
-2(x-1)^2\Bigr ]+2|D|^2\Bigl [\beta^2+(x-1)^2\Bigr ]}.
\end{eqnarray}
It should be noted that all of these asymmetries are odd under $CP$ but
even under $T$. Thus they require a non--zero phase difference between
the amplitudes $B$ and $D$, induced by final state interactions. This is
evident from the appearance of the factor $\mbox{Im}(B^*D)$ in the
asymmetric terms in Eq. (17), and in Eq. (20). The asymmetries are also
proportional to the  parameter $\xi$, which is approximately $0.16$ for
muon pairs.

\section{Energy Spectra and Correlations: Relation to Helicity Structure
of $H\to ZZ$ Amplitude}
Quite generally, the decay $H\to ZZ$ produces a system of two $Z$ bosons in
the helicity state
\begin{eqnarray}
|ZZ\rangle =c_+|++\rangle +c_-|--\rangle +c_0|00\rangle .
\end{eqnarray}
Couplings of the form $B$, $C$ and
$D$ (Eq. (6)) give rise to the following characteristic helicity
wave--functions:
\begin{eqnarray}
\begin{array}{c@{\quad : \quad}c}
Bg_{\mu\nu}\varepsilon_1^{\mu}\varepsilon_2^{\nu}&|++\rangle +|--\rangle +
\displaystyle\frac{1+\beta^2}{1-\beta^2}|00\rangle \\
\displaystyle\frac{C}{m_Z^2}p_{1\nu}p_{2\mu}\varepsilon_1^{\mu}
\varepsilon_2^{\nu}&|00\rangle \\
\displaystyle\frac{D}{m_Z^2}\varepsilon_{\mu\nu\rho\sigma}p_1^{\rho}
p_2^{\sigma}\varepsilon_1^{\mu}
\varepsilon_2^{\nu}&|++\rangle -|--\rangle .
\end{array}
\end{eqnarray}
Notice that the standard Higgs boson coupling generates a transversely
polarized state $|++\rangle +|--\rangle $ mixed with a specific amount of
longitudinal polarization $|00\rangle $. In the high energy limit $\beta\to
1$, the longitudinal component dominates. In comparison, a pseudoscalar Higgs
decays into a transversely polarized state $|++\rangle -|--\rangle $. A
scalar coupling of the form $\displaystyle\frac{C}{m_Z^2}p_{1\nu}p_{2\mu}
\varepsilon_1^{\mu}
\varepsilon_2^{\nu}$ generates a state of longitudinal polarization
$|00\rangle $ only.\\
\indent It is possible to relate the energy distributions derived in the
preceding Section to the helicity structure of the $H\to ZZ$ amplitude.
To this end, it is important to note (i) that, as far as the energy
distributions are concerned, the transverse and
longitudinal components of the $ZZ$ wave--function add
incoherently; (ii) that the energy spectra do not distinguish between the
states $|++\rangle +|--\rangle $ and $|++\rangle -|--\rangle $. In the
$CP$--invariant limit, therefore, the energy distribution of the final
fermion system can be written as follows:\\
(a) One--dimensional distribution
\begin{eqnarray}
\frac{1}{\Gamma}\frac{\mbox{d}\Gamma}{\mbox{d}x}=P_Tf_T(x)+P_Lf_L(x),
\end{eqnarray}
where $P_T$ and $P_L$ are the probabilities for transverse and longitudinal
polarization in the $ZZ$ wave--function ($P_T+P_L=1$), and $x$ is the
scaled energy ($x=4E/m_H$) of \it any \rm of the final fermions in $H\to
ZZ\to (f_1\bar{f_1})+(f_2\bar{f_2})$. The functions $f_T$ and $f_L$ are
given by
\begin{eqnarray}
f_T(x)&=&\frac{3}{8\beta^3}\Bigl [\beta^2+(x-1)^2\Bigr ],\nonumber\\
f_L(x)&=&\frac{3}{4\beta^3}\Bigl [\beta^2-(x-1)^2\Bigr ].
\end{eqnarray}
(b) Two--dimensional distribution
\begin{eqnarray}
\frac{1}{\Gamma}\frac{\mbox{d}\Gamma}{\mbox{d}x
\mbox{d}x'}
=P_T\Bigl [f_T(x)f_T(x')\pm\xi_1\xi_2g_T(x)g_T(x')\Bigr ]
+P_L\Bigl [f_L(x)f_L(x')\Bigr ],
\end{eqnarray}
with
\begin{eqnarray}
g_T(x)=\frac{3}{4\beta^2}(x-1).
\end{eqnarray}
In Eq. (25), the $+$ sign applies to a pair $(f_1,f_2)$ or $(\bar{f_1},
\bar{f_2})$, while the $-$ sign applies to a pair $(f_1,\bar{f_2})$.
It is easy to see that Eqs. (23) to (26) reproduce the
results (13) and (16), in the limit of $CP$--conservation.\\
\indent It is clear from the structure of Eq. (25) that, for $P_T\neq 0$, the
two--dimensional distribution is not simply a product of one--dimensional
spectra. The term proportional to $g_T(x)g_T(x')$ is indicative of the fact
that the transverse helicities of the two $Z$'s are \it correlated\rm . On
the other hand, for $P_T=0$, the two--particle distribution factorises
into a product of one--particle spectra. The three types of coupling
in Eq. (22) are obviously characterised by:
\begin{eqnarray}
\begin{array}{c@{\quad : \quad}c@{\quad , \quad}c}
B-\mbox{type}&P_T=\displaystyle\frac{2(1-\beta^2)^2}{3-2\beta^2+3\beta^4}&
P_L=\displaystyle\frac{(1+\beta^2)^2}{3-2\beta^2+3\beta^4}\\
C-\mbox{type}&P_T=0&P_L=1\\
D-\mbox{type}&P_T=1&P_L=0.
\end{array}
\end{eqnarray}

\section{Remarks on the Decay $H\to W^+W^-\to \mu^+\nu_{\mu}
\mu^-\bar{\nu_{\mu}}$}
Our results for the sequential decay $H\to ZZ\to \mu^+\mu^-\mu^+\mu^-$ are
immediately adaptable to the reaction
\begin{eqnarray}
H\to W^+W^-\to \mu^+\nu_{\mu}\mu^-\bar{\nu_{\mu}}.
\end{eqnarray}
One has only to put $\xi_{1,2}=1$ in Eqs. (13) or (16), and
interpret $\beta $ as $\sqrt{1-4m_W^2/m_H^2}$. If the $W^+W^-$
state is characterised by probabilities $P_T$ and $P_L$ for transverse
and longitudinal polarization, the one--particle spectrum of $\mu^+$ or
$\mu^-$ (in the $CP$--conserving limit) is \it identical \rm to that produced
by $H\to ZZ$, namely the spectrum given by Eq. (23). On the other hand the
correlated $\mu^+\mu^-$ distribution in reaction (28) differs significantly
from that in $H\to ZZ\to \mu^+\mu^-\mu^+\mu^-$, since it is obtained from
Eq. (25) with $\xi_1\xi_2 = 1$ instead of $\xi_1\xi_2 \approx 2.56\times
10^{-2}$. This gives
rise to a marked difference between the two reactions as illustrated in
Fig. 3, for the case of a standard Higgs boson, and in Fig. 4 for the case of
a pseudoscalar Higgs. Our results concerning the
energy distribution of the secondary leptons in $H\to W^+W^-\to l^+\nu_l
l^-\bar{\nu_l}$ coincide with those that we have reported in an earlier paper
\cite{arens}.\\
\newpage
\noindent Acknowledgements: We acknowledge useful discussions with Udo
Gieseler, and collaboration on a related paper. One of us (T.A.) is
recipient of a stipend from the state of Nordrhein Westfalen. The support
of the German Ministry of Research and Technology (BMFT) is acknowledged with
gratitude.

\newpage\Large
\noindent{\bf Figure Captions}
\normalsize
\begin{itemize}
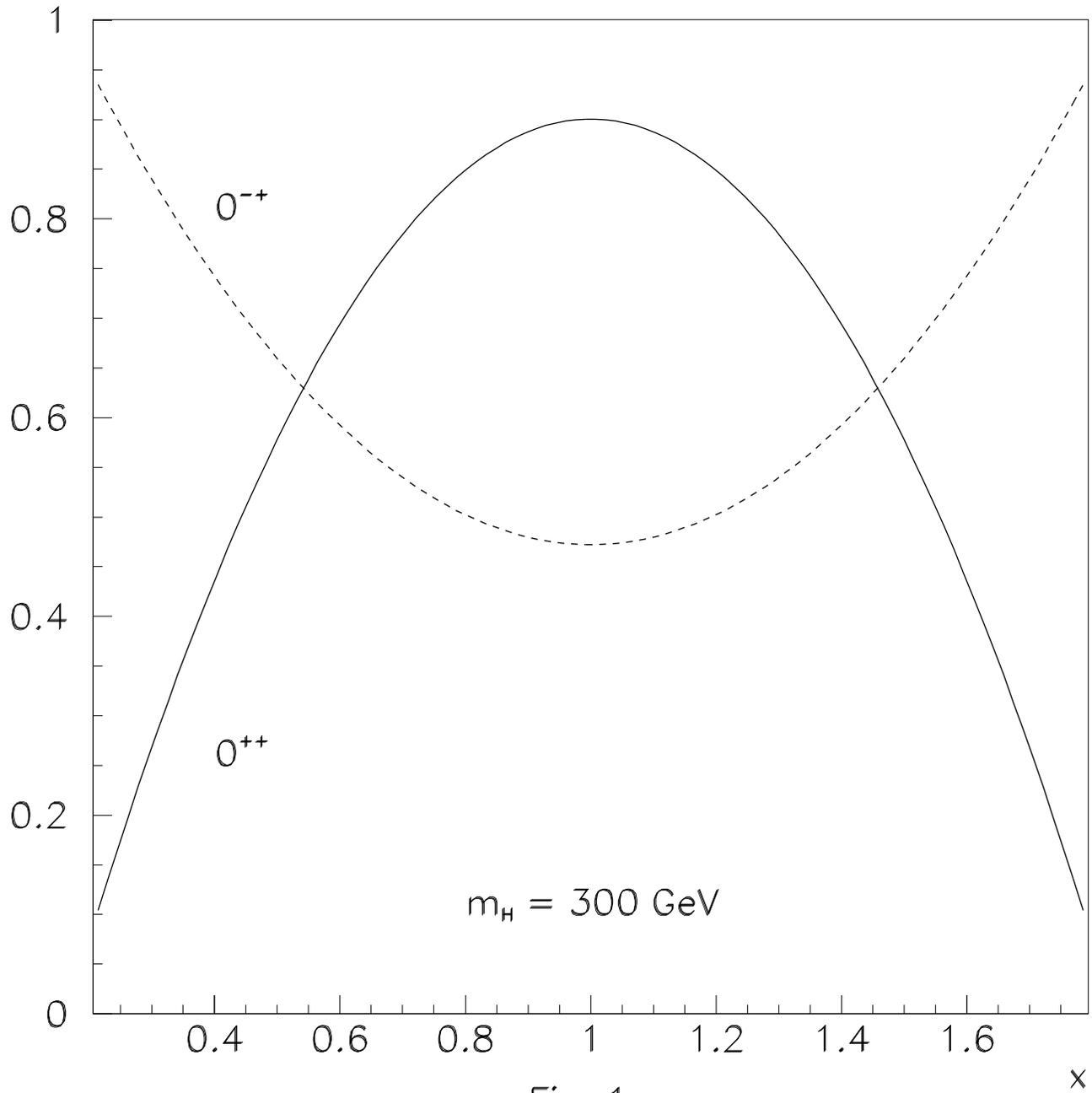
\item[Fig. 1.] Single particle energy spectra of a fermion $f$ in the
decay $H\to ZZ\to f+\cdots $. The full curve represents the scalar case
and the dashed curve the pseudoscalar case, for $m_H= 300 $ GeV.
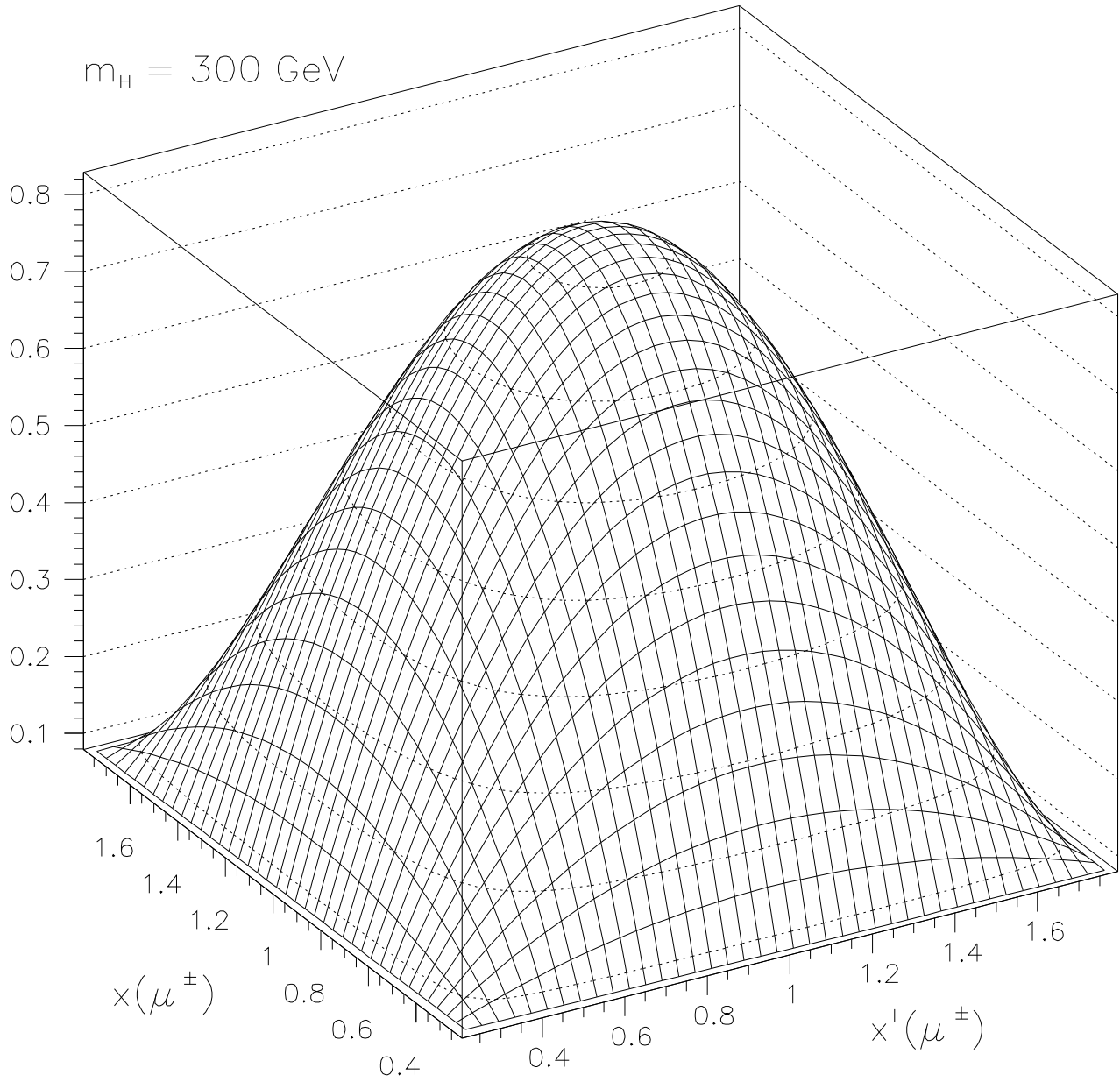
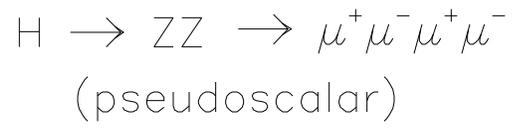
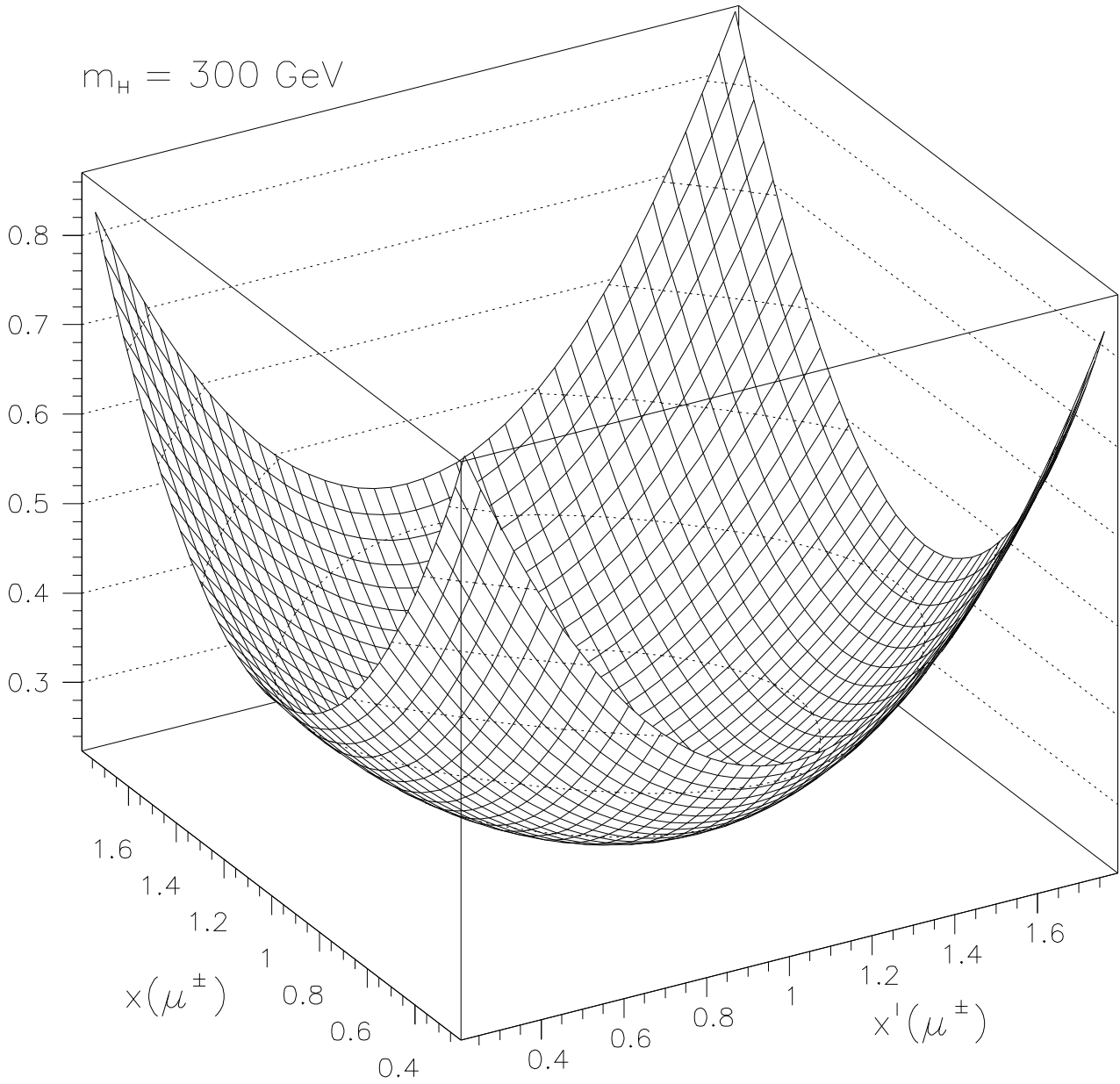
\item[Fig. 2.] Normalized energy distribution of like sign muon pairs
$\mu^+\mu^+$ or $\mu^-\mu^-$ in the decay $H\to ZZ\to\mu^+\mu^-\mu^+\mu^-$.
Fig. 2(a) shows the spectrum of a standard model Higgs boson, and Fig. 2(b)
that of a pseudoscalar Higgs, with $m_H= 300 $ GeV.
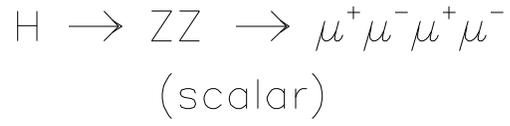
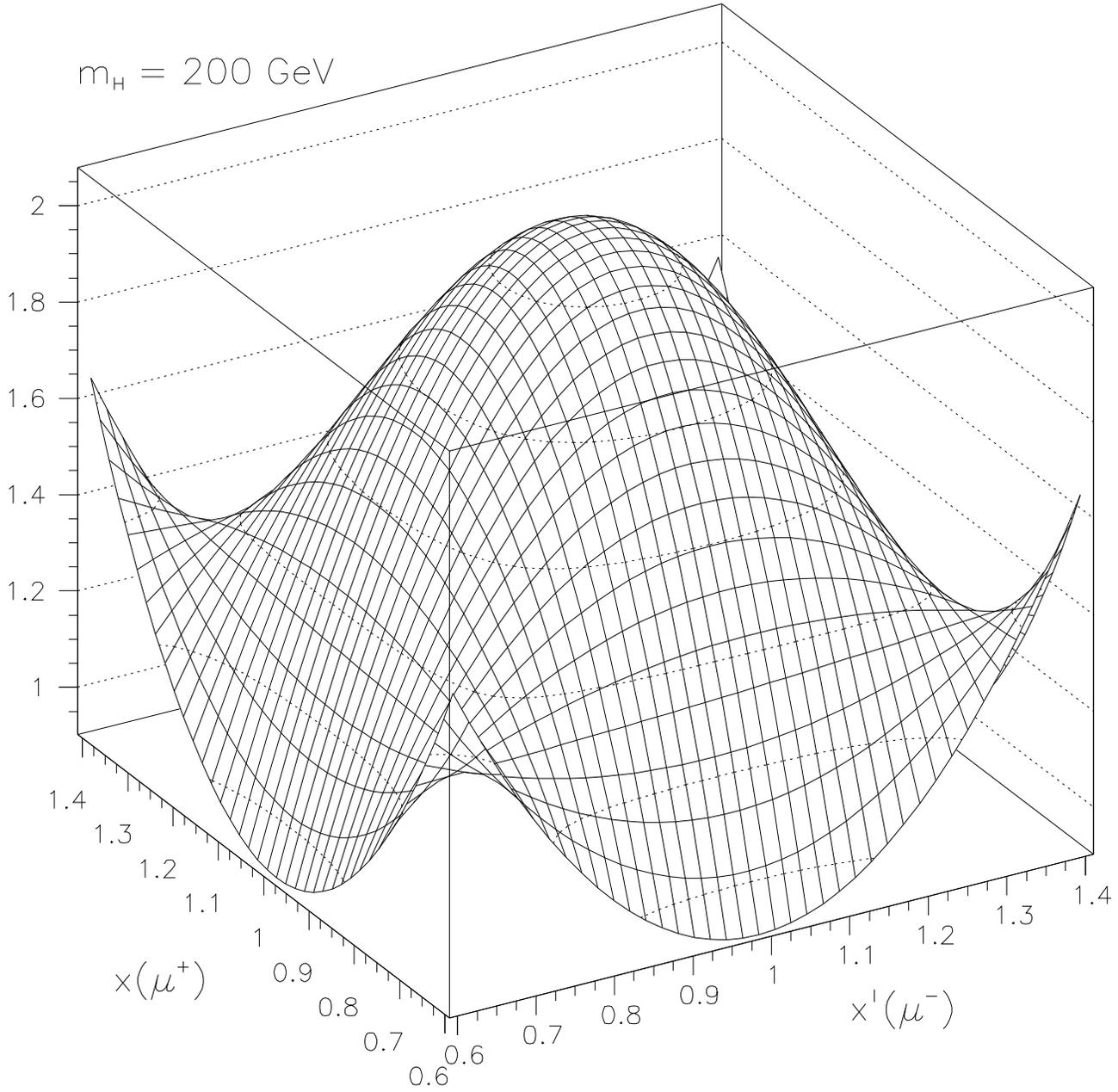
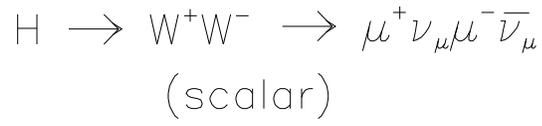
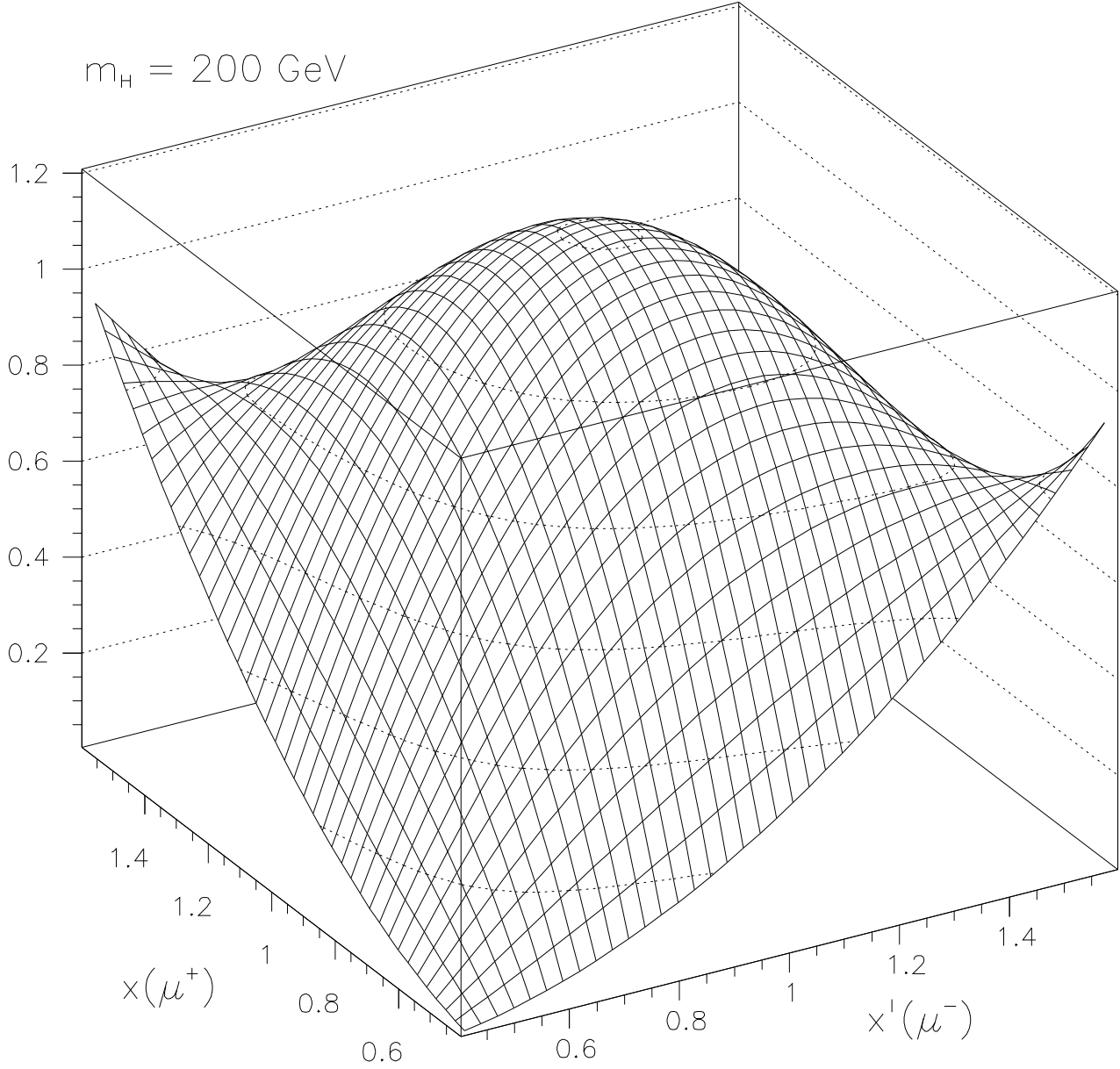
\item[Fig. 3.] Normalized energy distribution of unlike sign muon pairs
$\mu^+\mu^-$ in the decay of a standard model Higgs boson with mass
$m_H=200$ GeV. (a) $\mu^+\mu^-$ distribution in $H\to ZZ\to\mu^+\mu^-\mu^+
\mu^-$, with $\mu^+$ and $\mu^-$ chosen from different $Z$'s; (b)
$\mu^+\mu^-$ distribution in $H\to W^+W^-\to\mu^+\nu_{\mu}\mu^-
\bar{\nu_{\mu}}$.
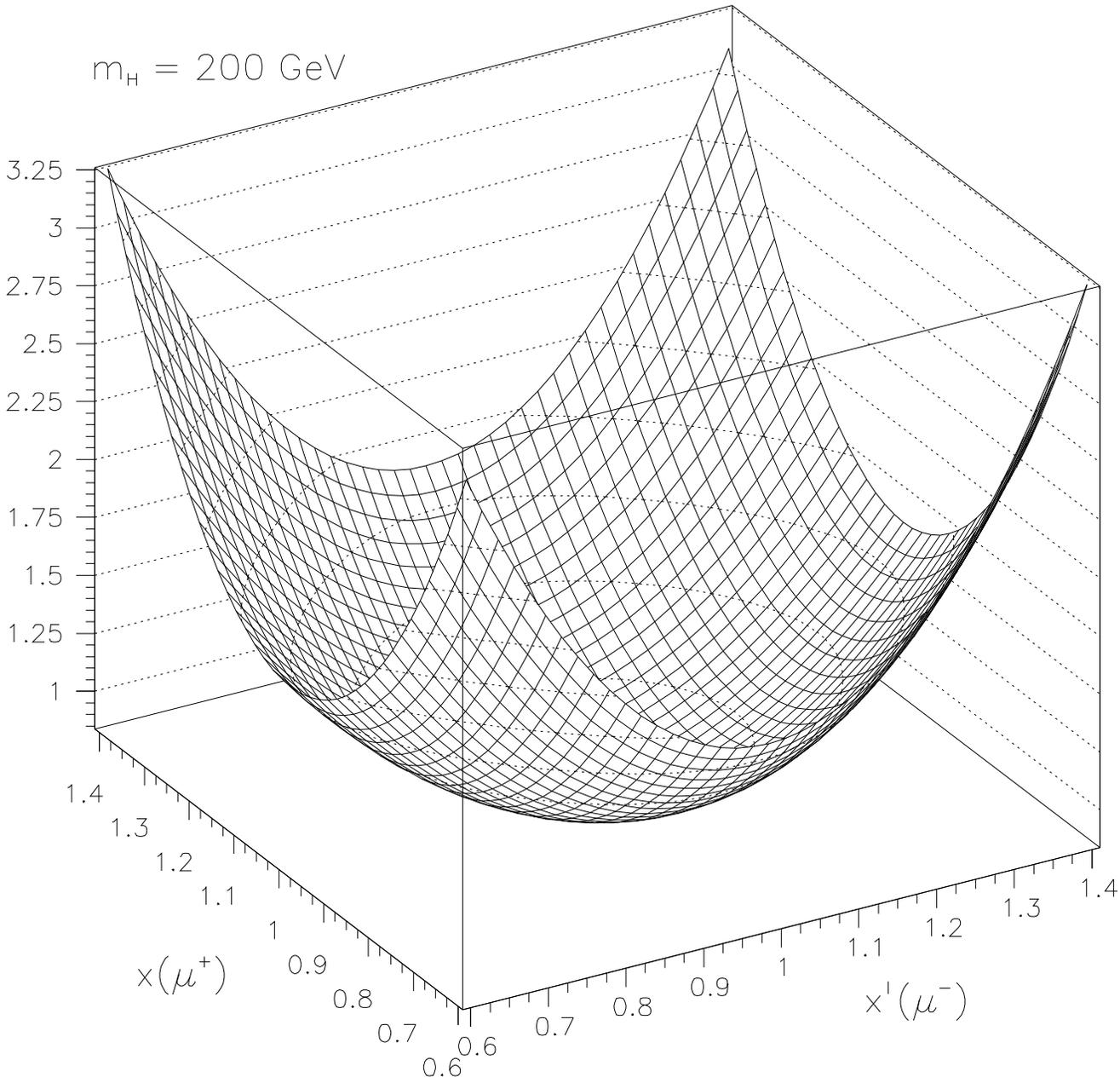
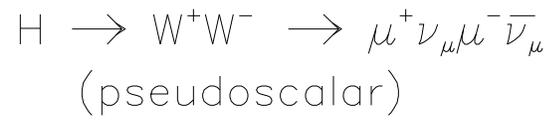
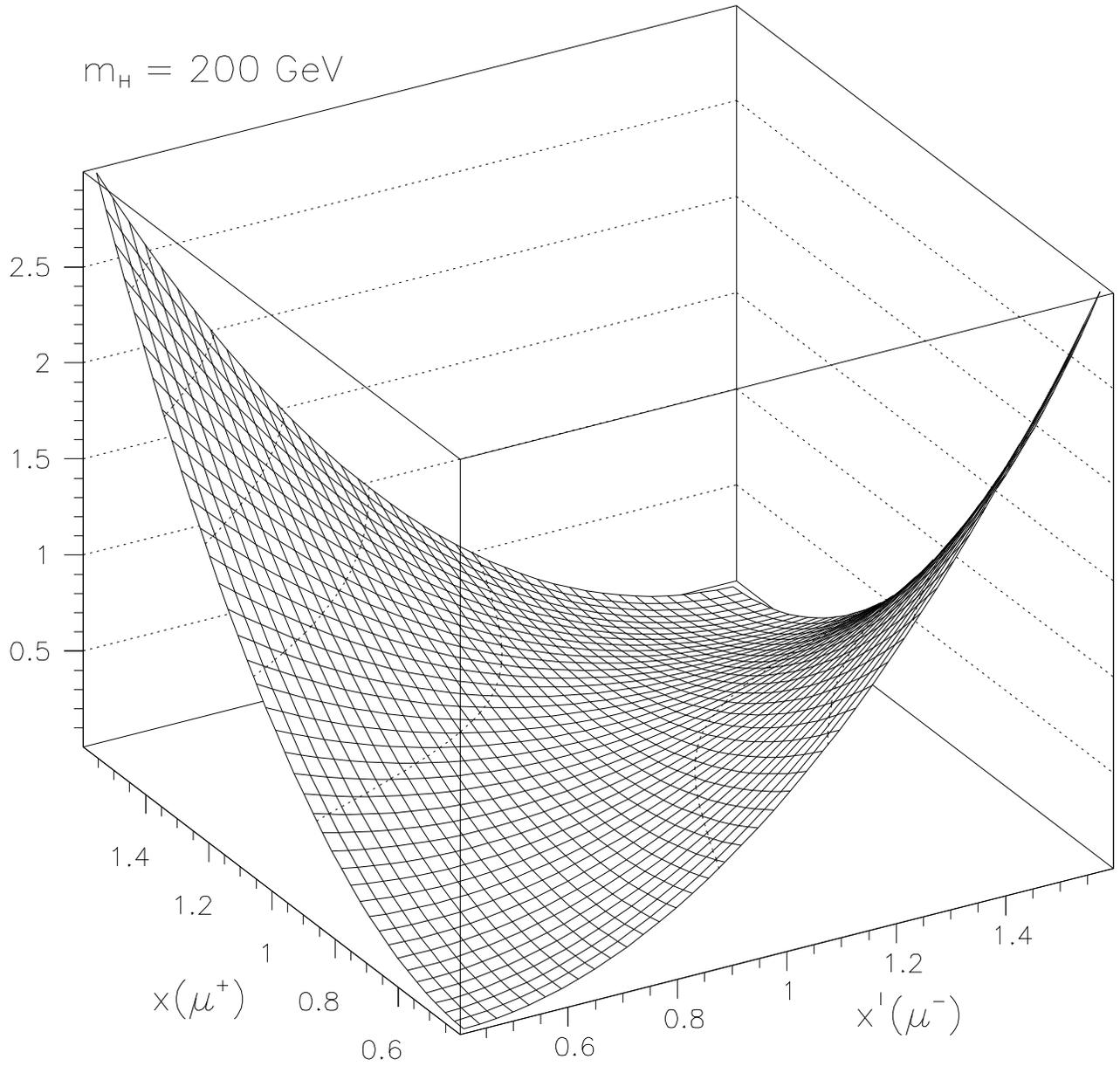
\item[Fig. 4.]  Normalized energy distribution of unlike sign
$\mu^+\mu^-$ pairs in the decay of a pseudoscalar Higgs boson
($m_H=200$ GeV). (a) $\mu^+\mu^-$ distribution in $H\to ZZ\to\mu^+\mu^-\mu^+
\mu^-$, with $\mu^+$ and $\mu^-$ chosen from different $Z$'s; (b)
$\mu^+\mu^-$ distribution in $H\to W^+W^-\to\mu^+\nu_{\mu}\mu^-
\bar{\nu_{\mu}}$.
\end{itemize}

\end{document}